# BATALIN-VILKOVISKY ALGEBRAS AND TWO-DIMENSIONAL TOPOLOGICAL FIELD THEORIES

E. GETZLER

Batalin-Vilkovisky algebras are a new type of algebraic structure on graded vector spaces, which first arose in the work of Batalin and Vilkovisky on gauge fixing in quantum field theory: a Batalin-Vilkovisky algebra is a differential graded commutative algebra together with an operator $\Delta : A_\bullet \to A_{\bullet+1}$ such that $\Delta^2 = 0$, and

$$\Delta(abc) = \Delta(ab)c + (-1)^{|a|}a\Delta(bc) + (-1)^{(|a|-1)|b|}b\Delta(ac)$$
$$- (\Delta a)bc - (-1)^{|a|}a(\Delta b)c - (-1)^{|a|+|b|}ab(\Delta c).$$

(Some references for Batalin-Vilkovisky algebras, with diverse applications to physics, are Schwarz [20], Witten [22] and Zwiebach [24].)

In this article, we show that there is a natural structure of a Batalin-Vilkovisky algebra on the cohomology of a topological conformal field theory in two dimensions. Lian and Zuckerman [16] have constructed a Batalin-Vilkovisky structure on the cohomology of a topological chiral field theory, and calculated it explicitly in the case of $D = 2$ string theory. Our approach to the study of this structure is quite different from that of Lian and Zuckerman, and makes use of a technique from algebraic topology: the theory of operads. (It was lectures of Maxim Kontsevitch at Harvard which first suggested a role for operads in topological field theory.) As a result, we obtain the stronger result that no additional relations on the resulting algebraic structure may be inferred from the axioms of a topological conformal field theory in restricted to genus 0 Riemann surfaces.

Since a topological conformal field theory has both chiral and anti-chiral sectors, it actually has not one but two Batalin-Vilkovisky structures, with a common product but two commuting operators $\Delta$ and $\bar\Delta$. (There are some similarities to the theory of bisymplectic structures.) The Batalin-Vilkovisky structure which we construct is the diagonal structure $\Delta - \bar\Delta$. In fact, the operators $\Delta$ and $\bar\Delta$ could also be obtained by the methods of this paper, by replacing the de Rham complex of our operad by its Dolbeault complex; indeed, our operad is a Stein manifold, and hence its de Rham and Dolbeault cohomology are isomorphic. Furrther, our methods actually construct, at least heuristically, a Batalin-Vilkovisky structure for any topological field theory in two dimensions.

The author is partially supported by a fellowship of the Sloan Foundation and a research grant of the NSF.

We will adopt a definition of a topological conformal field theory inspired by the operator formalism of bosonic string theory, in which the functional integrals are differential forms on Segal's category [21]. Let $\widehat{\mathcal{M}}_{g,n}$ denote the moduli space of connected Riemann surfaces of genus $g$, together with a biholomorphic map from the disjoint union $\coprod_{i=1}^{n} D$ of $n$ discs, such that the images of the discs are disjoint. For example, $\widehat{\mathcal{M}}_{0,n}$ is the space of all biholomorphic maps from $\coprod_{i=1}^{n} D$ to $S^2$ such that images of the discs are disjoint, modulo the action of the group $\mathrm{PSL}(2,\mathbb{C})$ on $S^2$.

In a conformal field theory, the functional integral is a function on $\widehat{\mathcal{M}}_{g,n}$ with values in $\mathcal{E}^{(n)}$, the $n$-th tensor power of the Hilbert space $\mathcal{E}$. In a topological conformal field theory, this is modified in the following way: now $\mathcal{E}$ is a complex with differential $Q$, the functional integral is a differential form on $\widehat{\mathcal{M}}_{g,n}$ with values in $\mathcal{E}^{(n)}$, and the functional integral is closed with respect to the total differential $d + Q$. Witten's formulas $[Q, S] = 0$, and
$$T_{\mu\nu} = \frac{\partial S}{\partial g^{\mu\nu}} = [Q, G_{\mu\nu}]$$
relating the degree zero component of the functional integral $S$ and the coefficients of its degree one component $G_{\mu\nu}$ (see [23]) are interpreted as the lowest two terms of the formula
$$(d+Q)(S - G_{\mu\nu}dg^{\mu\nu} + \dots) = 0$$
for the full functional integral $\omega = S - G_{\mu\nu}dg^{\mu\nu} + \dots$. To complete the definition of a topological conformal field theory, we impose sewing and equivariance axioms; this is explained in Section 3, following Segal. (The author is very grateful to Graeme Segal for teaching him this formalism.)

Our method may be understood by means of an analogy. If $M$ is a based topological space, let $\Omega^2 M$ be the double loop space
$$\Omega^2 M = \{f : \mathbb{C} \to M \mid f(z) = * \text{ for } |z| \geq 1\}.$$
The circle acts on $\Omega^2 M$ by rotating the complex plane $\mathbb{C}$. It may be proved using operads that the homology of $\Omega^2 M$ is a Batalin-Vilkovisky algebra: the product is the well-known Pontryagin product, induced by the product on double loops $\Omega^2 M \times \Omega^2 M \to \Omega^2 M$ defined by the formula
$$(fg)(z) = \begin{cases} f(2z+1), & \mathrm{Re}(z) < 0, \\ g(2z-1), & \mathrm{Re}(z) > 0, \end{cases}$$
and the operator $\Delta x = e_*([S^1] \times x)$ is defined using the circle action $e : S^1 \times \Omega^2 M \to \Omega^2 M$.

The resulting analogy between the cohomology of a topological conformal field theory and the homology of a double loop space is not perfect, since the cohomology of a topological conformal field theory has an extra piece of structure, a linear map $\int : \mathcal{H} \to \mathbb{C}$ such that the quadratic form

$$\langle x, y \rangle = \int xy$$

is non-degenerate, and which satisfies the formula

$$\int (\Delta x) y = (-1)^{|x|} \int x (\Delta y).$$

In this respect, the cohomology a topological conformal field theory is reminiscent of the algebra of differential forms $\Omega^\bullet(M)$ on a compact oriented manifold $M$ with a differentiable circle action, in which case the role of the $\Delta$ is played by the operator

$$D = \pi_* e^* : \Omega^\bullet(M) \to \Omega^{\bullet-1}(M);$$

here $e : S^1 \times M \to M$ is the circle action and $\pi : S^1 \times M \to M$ is the projection onto the second factor. On the other hand, since the operator $D$ induces a derivation of the product on the de Rham cohomology $H^\bullet(M)$, unlike in a general Batalin-Vilkovisky algebra.

Let us summarize these analogies with our version of the Rosetta Stone:

| topological field theory | double loop space | manifold with circle action |
|---|---|---|
| Hilbert space $\mathcal{E}$ | $C_\bullet(\Omega^2 M)$ | $\Omega^\bullet(M)$ |
| BRST operator $Q$ | boundary $\delta$ | exterior differential $d$ |
| physical Hilbert space: | singular homology: | de Rham cohomology: |
| $\mathcal{H} = H_\bullet(\mathcal{E}, Q)$ | $H_\bullet(\Omega^2 M)$ | $H^\bullet(M)$ |
| product, $\Delta$ | Pontryagin product, $\Delta$ | wedge product, $D$ |

In each column of this table, we could as well take the equivariant (co)homology with respect to the circle action. In another article, we will discuss the analogy between equivariant cohomology and the coupling of a topological conformal field theory to topological gravity.

For a topological conformal field theory obtained by the twisting operation of Witten and Eguchi-Yang from a unitary $N = 2$ supersymmetric conformal field theory, the Batalin-Vilkovisky algebra which we obtain is naturally isomorphic to the chiral primary ring of the original supersymmetric field theory (see Lerche-Vafa-Warner [15] and Dijkgraaf-Verlinde-Verlinde [6] for thorough discussions of this ring). This is a rather degenerate example of a Batalin-Vilkovisky algebra, since the operator $\Delta$ vanishes; in this, it is analogous to a trivial action of the circle on a manifold. Other examples of topological field theoriesare provided by (bosonic) string theories, which tend to have infinite-dimensional cohomology and non-vanishing $\Delta$.

The author was led to the results of this article through his joint work with John Jones [10], where we show that the braid operad (defined in Section 4) satisfies a certain self-duality property. Independently, Ginzburg and Kapranov [11] have introduced the notion of a Koszul operad, and shown that the associative, commutative and Lie operads are Koszul. (This is an analogue of the notion of a Koszul algebra, that is, an algebra with quadratic relations whose Koszul complex, as defined by Priddy [19], is acyclic: examples of Koszul algebras are the universal enveloping algebra of a Lie algebra and the Steenrod algebra.) Implicit in Theorem 4.5 is the statement that the Batalin-Vikovisky operad is Koszul.

This article is a revised version of a preprint of December, 1992. In preparing this revision, he has taken into account remarks of Greg Moore and Gregg Zuckerman, which were helpful in understanding the role of chirality, and the relationship with the work of Lian-Zuckerman. He thanks Barton Zwiebach for stimulating conversations on Batalin-Vilkovisky algebras, and for correcting the definition of a trace contained in the earlier version, and Larry Breen for a number of helpful remarks. Since the original preprint, some related papers have appeared: Hořava [13] and Penkava-Schwartz [18].

## 1. Batalin-Vilkovisky algebras

By a chain complex, we mean a vector space over $\mathbb{C}$ graded by integers with differential $Q$ lowering degree by 1. Eventually, we will specialize to the case where $V$ is the space of states (including ghosts) of a topological field theory in two dimensions. The operator $Q$ is known in that setting as the BRST operator. Our convention on grading differs from the usual convention of topological field theory, in which $Q$ raises degree by 1: our ghost number is the negative of the usual one.

If $V$ is a graded vector space, we denote by $\Sigma^n V$ the $n$-th suspension of $V$, which is the graded vector space such that

$$(\Sigma^n V)_i = V_{i-n}.$$

The $k$-fold graded tensor product $V^{(k)}$ carries an action of the symmetric group $\mathbb{S}_k$, such that the transposition $s_i$ which exchanges $i$ and $i+1$ acts by

$$s_i(v_1 \otimes \ldots \otimes v_k) = (-1)^{|v_i||v_{i+1}|} v_1 \otimes \ldots \otimes v_{i+1} \otimes v_i \otimes \ldots \otimes v_k.$$

A Lie bracket of degree $m$ on a chain complex $V$ is a Lie bracket on $\Sigma^m V$, that is, a bilinear map $[-,-] : V \otimes V \to V$ such that $[v,w] = -(-1)^{(|v|-m)(|w|-m)}[w,v]$, and which satisfies the Jacobi rule

$$[u,[v,w]] = [[u,v],w] + (-1)^{(|u|-m)(|v|-m)}[v,[u,w]].$$

**Definition 1.1.** A braid algebra is a differential graded commutative algebra together with a Lie bracket of degree 1 satisfying the Poisson relation

$$[u,vw] = [u,v]w + (-1)^{|u|(|v|-1)} v[u,w].$$

An identity in a braid algebra is an element 1 of degree 0 which is an identity for the product and such that $[1, -] = 0$.

A Batalin-Vilkovisky algebra $A$ is a differential graded commutative algebra together with an operator $\Delta : A_\bullet \to A_{\bullet+1}$ such that $\Delta^2 = 0$, and

$$\Delta(abc) = \Delta(ab)c + (-1)^{|a|}a\Delta(bc) + (-1)^{(|a|-1)|b|}b\Delta(ac)$$
$$- (\Delta a)bc - (-1)^{|a|}a(\Delta b)c - (-1)^{|a|+|b|}ab(\Delta c).$$

An identity in a Batalin-Vilkovisky algebra is an element 1 of degree 0 which is an identity for the product and such that $\Delta 1 = 0$. (All of the Batalin-Vilkovisky algebras which we discuss will possess an identity.)

We now show that a Batalin-Vilkovisky algebra is a special type of braid algebra.

**Proposition 1.2.** *A Batalin-Vilkovisky algebra is a braid algebra equipped with an operator $\Delta : A_\bullet \to A_{\bullet+1}$, such that $\Delta^2 = 0$, such that the bracket and $\Delta$ are related by the formula*

$$[a, b] = (-1)^{|a|}\Delta(ab) - (-1)^{|a|}(\Delta a)b - a(\Delta b).$$

*Furthermore, in a Batalin-Vilkovisky algebra, $\Delta$ satisfies the formula*

$$\Delta[a, b] = [\Delta a, b] + (-1)^{|a|-1}[a, \Delta b].$$

*Proof.* Suppose $A_\bullet$ is a Batalin-Vilkovisky algebra, and the Poisson bracket $[a, b]$ is defined as above. We check that $[a, b]$ is antisymmetric:

$$[a, b] = (-1)^{|a|}\Delta(ab) - (-1)^{|a|}(\Delta a)b - a(\Delta b)$$
$$= (-1)^{|a|+|a||b|}\Delta(ba) - (-1)^{|a|+(|a|-1)|b|}b(\Delta a) - (-1)^{|a|(|b|-1)}(\Delta a)b$$
$$= (-1)^{|a||b|+|a|+|b|}[b, a].$$

Next we verify the Leibniz rule for $\Delta[a, b]$. On the one hand,

$$\Delta[a, b] = (-1)^{|a|}\Delta(\Delta(ab)) - (-1)^{|a|}\Delta((\Delta a)b) - \Delta(a(\Delta b)),$$

while on the other,

$$[\Delta a, b] + (-1)^{|a|-1}[a, \Delta b] = (-1)^{|a|-1}\Delta((\Delta a)b) - (-1)^{|a|-1}(\Delta^2 a)b - (\Delta a)(\Delta b)$$
$$- \Delta(a(\Delta b)) + (\Delta a)(\Delta b) + (-1)^{|a|-1}a(\Delta^2 b)$$
$$= (-1)^{|a|-1}\Delta((\Delta a)b) - \Delta(a(\Delta b)), \quad \text{since } \Delta^2 = 0.$$

Next we check that the Poisson relation for a Poisson bracket is equivalent to the identity for $\Delta(abc)$ in a Batalin-Vilkovisky algebra:

$$\begin{aligned}
[a, bc] &- [a, b]c - (-1)^{(|a|-1)|b|}b[a, c] \\
&= (-1)^{|a|}\Delta(abc) - (-1)^{|a|}(\Delta a)bc - a\Delta(bc) \\
&\quad - (-1)^{|a|}\Delta(ab)c + (-1)^{|a|}(\Delta a)bc + a(\Delta b)c \\
&\quad - (-1)^{(|a|-1)|b|}\Big((-1)^{|a|}b\Delta(ac) - (-1)^{|a|}b(\Delta a)c - ba(\Delta c)\Big).
\end{aligned}$$

After cancelling two of the terms proportional to $(\Delta a)bc$, we obtain the identity for $(-1)^{|a|}\Delta(abc)$ in a Batalin-Vilkovisky algebra.

Finally, we turn to the Jacobi rule. We have

$$\begin{aligned}
[[x, a], b] &+ (-1)^{(|x|-1)(|a|-1)}[a, [x, b]] \\
&= (-1)^{|x|+|a|-1}\Delta([x, a]b) - (-1)^{|x|+|a|-1}(\Delta[x, a])b - [x, a](\Delta b) \\
&\quad + (-1)^{(|x|-1)(|a|-1)}\Big((-1)^{|a|}\Delta(a[x, b]) - (-1)^{|a|}(\Delta a)[x, b] - a(\Delta[x, b])\Big) \\
&= (-1)^{|x|+|a|-1}\Delta([x, ab]) - (-1)^{|x|+|a|-1}[\Delta x, a]b - (-1)^{(|x|-1)(|a|-1)}a[\Delta x, b] \\
&\quad - (-1)^{|a|}[x, \Delta a]b - (-1)^{(|x|-1)(|a|-1)+|a|}(\Delta a)[x, b] \\
&\quad - [x, a](\Delta b) - (-1)^{(|x|-1)(|a|-1)+|a|-1}a[x, \Delta b], \\
&= (-1)^{|a|}[x, \Delta(ab)] - (-1)^{|a|}[x, (\Delta a)b] - [x, a(\Delta b)] = [x, [a, b]].
\end{aligned}$$

where we have used the Poisson rule for $[x, ab]$, $[x, (\Delta a)b]$ and $[x, a(\Delta b)]$ and the Leibniz rule for $\Delta[x, a]$ and $\Delta[x, b]$. $\square$

Lian and Zuckerman employ the term Gerstenhaber algebra for what we call a braid algebra. Gerstenhaber showed that the Hochschild cohomology $H^\bullet(A, A)$ of a graded algebra is what we call a braid algebra [9]. This cohomology is defined using the Hochschild cochains, which are the multilinear maps from $A$ to itself:

$$C^\bullet(A, A) = \sum_{k=0}^\infty \operatorname{Hom}(A^{(k)}, A).$$

Gerstenhaber defines an operation $c_1 \circ c_2$ on $C^\bullet(A, A)$ by the formula

$$\begin{aligned}
(c_1 \circ c_2)&(a_1, \ldots, a_n) \\
&= \sum_{0 \le i \le j \le n} (-1)^{(|c_2|-1)(|a_1|+\cdots+|a_i|-i|)} c_1(a_1, \ldots, a_i, c_2(a_{i+1}, \ldots, a_j), a_{j+1}, \ldots, a_n).
\end{aligned}$$

The bracket

$$[c_1, c_2] = c_1 \circ c_2 - (-1)^{(|c_1|-1)(|c_2|-1)} c_2 \circ c_1$$

gives $\Sigma C^\bullet(A, A)$ the structure of a Lie algebra. Note that this structure does not depend on the product on $A$. The Hochschild differential is the operation $[m, -]$, where $m$ is the cochain defined by the product on $A$,
$$m(a_1, a_2) = (-1)^{|a_1|} a_1 a_2.$$
The equation $[m, [m, -]] = 0$ is equivalent to the associativity of $A$, which may be expressed by the formula $m \circ m = 0$.

The cup product on $C^\bullet(A, A)$ is given by the formula
$$(c_1 \cup c_2)(a_1, \ldots, a_n) = \sum_{i=0}^{n} c_1(a_1, \ldots, a_i) \, c_2(a_{i+1}, \ldots, a_n).$$
It is associative, but not commutative unless $A$ is, nor does it satisfy the Poisson relation with respect to the bracket. However, on taking cohomology, the relations of a braid algebra are satisfied.

In the special case that $A$ is the algebra of differentiable functions $C^\infty(M)$ on a manifold $M$, the Hochschild cohomology was shown by Hochschild-Kostant-Rosenberg [12] to be naturally isomorphic to the space of multivectors $\Gamma(M, \bigwedge^\bullet TM)$. With this identification, the cup product may be identified with the wedge product on $\Gamma(M, \bigwedge^\bullet TM)$, while the Gerstenhaber bracket may be identified with the Schouten-Nijenhuis-Richardson bracket.

To give some experience with these algebraic structures, let us write down explicitly the free braid and Batalin-Vilkovisky algebras on a single generator $x$ of degree $n$.

(1) If $x$ is odd, the free braid algebra on $x$ is equal to the free graded commutative algebra on $x$ (that is, the exterior algebra on $x$), and has vanishing bracket.

(2) If $n$ is even, the free braid algebra generated by $x$ has as its underlying graded commutative algebra the free graded commutative algebra with generators $x$ and $[x, x]$ (of degrees $n$ and $2n + 1$), and the bracket is determined by the Poisson and Jacobi relations.

(3) The free Batalin-Vilkovisky on $x$ has as its underlying graded commutative algebra the free graded commutative algebra on $x$, $\Delta x$ and $\Delta(x^2)$ (of degrees $n$, $n+1$ and $2n+1$ respectively).

Batalin-Vilkovisky algebras arise naturally in the theory of double-loop spaces. If $X$ is a based topological space, the homology of the based loop space
$$\Omega M = \{\gamma : \mathbb{R} \to M \mid \gamma(x) = * \text{ for } |x| \geq 1\}$$
has a natural structure of an associative algebra, with respect to the Pontryagin product, or composition of loops. (In this article, all homology groups $H_\bullet(M)$ will be taken over the complex numbers $\mathbb{C}$.) Since the double loop space $\Omega^2 M$ is homeomorphic to the loop space of $\Omega M$, we see that the homology of $\Omega^2 M$ is an associative algebra. Indeed, since $\Omega M$ is an H-space, the product on $H_\bullet(\Omega^2 M)$ is graded commutative.

An action of the circle on a based topological space $X$, preserving the base-point, induces an operator $\Delta : H_\bullet(X) \to H_{\bullet+1}(X)$ on the homology of $X$, the push-forward $\Delta x = e_*([S^1] \times x)$, where $e : S^1 \times X \to X$ is the multiplication map and $[S^1] \in H_1(S^1)$ is the fundamental class of the circle.

If the circle group acts on a based space $M$, it acts on $\Omega^2 M$ by rotating the sphere $S^2$, and simultaneously rotating the space $M$ by its underlying circle action, and thus induces an operator

$$\Delta : H_\bullet(\Omega^2 M) \to H_{\bullet+1}(\Omega^2 M).$$

We will prove in Section 4 that the Pontryagin product and the operator $\Delta$ define on $H_\bullet(\Omega^2 M)$ a Batalin-Vilkovisky algebra: the associated bracket is known as the Browder operation.

Let us illustrate this in the case where $M$ is a suspension $\Sigma^2 X$, where $X$ is connected and carries a circle action preserving the base-point. By Cohen's results [5], the homology of $\Omega^2 \Sigma^2 X$ is the free braid algebra on $\tilde{H}_\bullet(X)$, the reduced homology of $X$. The action of $\Delta$ on $H_\bullet(\Omega^2 \Sigma^2 X)$ is determined by its action on $\tilde{H}_\bullet(X)$, together with the formula

$$\Delta(xy) = (\Delta x)y + (-1)^{|x|} x(\Delta y) + (-1)^{|x|}[x, y].$$

For example, if $f$ is a polynomial and $x \in H_\bullet(X)$ has even degree, then we have the formula

$$\Delta(f(x)) = f'(x)\Delta x + \frac{1}{2} f''(x)[x, x] \in H_\bullet(\Omega^2 \Sigma^2 X).$$

## 2. Odd symplectic geometry

In the work of Batalin and Vilkovisky, Batalin-Vilkovisky algebras arise as the algebra of functions on an odd symplectic supermanifold. In this section, we recall their construction: for more details, see Witten [22] and Schwarz [20].

Let $M$ be an odd symplectic supermanifold, that is, a supermanifold with a closed, nondegenerate two-form $\omega$ of odd parity. To a function $f \in C^\infty(M)$, we associate a Hamiltonian vector field $H_f$ by the formula $\iota(H_f)\omega = df$. With the Poisson bracket

$$[f, g] = (-1)^{|f|-1} H_f(g),$$

$C^\infty(M)$ is a braid algebra, as we will now show. (In this example, the braid algebra is only $\mathbb{Z}/2$-graded and not $\mathbb{Z}$-graded.)

**Proposition 2.1.** *Let $M$ be an odd symplectic supermanifold. For $f, g, h \in C^\infty(M)$,*

(1) $\mathcal{L}(H_f)\omega = 0$
(2) $H_{fg} = H_f g + (-1)^{|f|} f H_g$
(3) $[H_f, H_g] = H_{[f,g]}$
(4) $[f, gh] = [f, g]h + (-1)^{(|f|-1)|g|} g[f, h]$
(5) $[f, [g, h]] = [[f, g], h] + (-1)^{(|f|-1)(|g|-1)}[g, [f, h]]$

*Proof.* Part (1) follows from the formula

$$\mathcal{L}(H_f)\omega = \iota(H_f)d\omega + d(\iota(H_f)\omega)$$

The first term vanishes, since $d\omega = 0$, while the second is proportional to $d^2 f$, and hence vanishes too.

Part (2) follows from the calculation

$$\begin{aligned}\iota(H_{fg})\omega &= d(fg) = (df)g + (-1)^{|f|}f(dg) \\ &= \iota(H_f)\omega g + (-1)^{|f|}f\iota(H_g)\omega \\ &= \iota(H_f g + (-1)^{|f|}fH_g)\omega.\end{aligned}$$

Part (3) is shown as follows:

$$\begin{aligned}\iota([H_f, H_g])\omega &= \mathcal{L}(H_f)\iota(H_g)\omega, \quad \text{since } \mathcal{L}(H_f)\omega = 0, \\ &= \mathcal{L}(H_f)dg \\ &= (-1)^{|f|-1}d(H_f(g)) \\ &= (-1)^{|f|-1}\iota\bigl(H_{H_f(g)}\bigr)\omega = \iota\bigl(H_{[f,g]}\bigr)\omega.\end{aligned}$$

The Poisson relation follows from the formula

$$H_f(gh) = H_f(g)h + (-1)^{(|f|-1)|g|}gH_f(h).$$

The Jacobi rule is shown as follows:

$$\begin{aligned}[f, [g, h]] &= (-1)^{|f|+|g|}H_f(H_g(h)) \\ &= (-1)^{|f|+|g|}[H_f, H_g](h) + (-1)^{(|f|-1)(|g|-1)+|f|+|g|}H_g(H_f(h)) \\ &= (-1)^{|f|+|g|}H_{[f,g]}(h) + (-1)^{(|f|-1)(|g|-1)}[g, [f, h]] \\ &= [[f, g], h] + (-1)^{(|f|-1)(|g|-1)}[g, [f, h]]. \quad \square\end{aligned}$$

Let $M$ be an odd symplectic supermanifold, and let $\mu$ be a nowhere-vanishing section $\mu$ of the Berezinian bundle $\text{Ber}(M)$ of $M$: this is the line bundle over $M$, generalizing the bundles of densities on a manifold, and characterized by the existence of an integral on sections of compact support $\int : \Gamma_c(M, \text{Ber}(M)) \to \mathbb{C}$. (The support of a section is a closed subset of the manifold underlying $M$.) The section $\mu$ induces an integral on functions with compact support on $M$,

$$f \in C_c^\infty(M) \mapsto \int_\mu f.$$

With respect to this integral, one defines a divergence operator $\text{div}_\mu X$ mapping vector fields on $M$ to functions, characterized by the formula

$$\int_\mu (\text{div}_\mu X)f = -\int_\mu X(f).$$

**Lemma 2.2.** (1) *If $X$ is a vector field, let $X^* = -X - \operatorname{div}_\mu X$. Then*

$$\int_\mu f\, X(g) = (-1)^{|f||X|} \int_\mu X^*(f)\, g.$$

(2) $\operatorname{div}_\mu(fX) = f \operatorname{div}_\mu X - (-1)^{|f||X|} X(f)$

(3) *If $S$ is an even function on $M$, then $\operatorname{div}_{\exp(S)\mu} X = \operatorname{div}_\mu X + X(S)$.*

*Proof.* To prove Part (1), note that

$$\int_\mu X(f)\, g + (-1)^{|f||X|} \int_\mu f\, X(g) = \int_\mu X(fg) = -\int_\mu (\operatorname{div}_\mu X)\, fg.$$

To prove Part (2), observe that by Part (1),

$$\begin{aligned}(fX)^* g &= (-1)^{|f||X|} X^*(fg) \\ &= fX^*(g) - (-1)^{|f||X|} X(f)g.\end{aligned}$$

To prove Part (3), we observe that

$$\begin{aligned}\int_\mu (\operatorname{div}_{\exp(S)\mu} X) f \exp(S) &= -\int_\mu X(f) \exp(S) \\ &= \int_\mu (\operatorname{div}_\mu X) f \exp(S) + (-1)^{|X||f|} \int_\mu f\, X(S) \exp(S). \quad \square\end{aligned}$$

Let $\Delta$ be the odd operator on functions on $M$ defined by the formula

$$\Delta f = \operatorname{div}_\mu H_f.$$

This operator measures the extent to which the Hamiltonian vector field $H_f$ fails to preserve the Berezinian $\mu$. No choice of Berezinian $\mu$ can lead to a vanishing operator $\Delta$: this is shown by the fact that $\Delta$ is a second-order differential operator with symbol the inverse of the symplectic form $\omega$. This is in contrast to even symplectic geometry, where Liouville's theorem shows that Hamiltonian vector fields preserve the Liouville measure.

A Batalin-Vilkovisky supermanifold $(M, \omega, \mu)$ is an odd symplectic supermanifold with Berezinian $\mu$, such that $\Delta^2 = 0$.

**Proposition 2.3.** *Let $(M, \omega, \mu)$ be a Batalin-Vilkovisky supermanifold.*

(1) *The algebra of functions $C^\infty(M)$ is a Batalin-Vilkovisky algebra.*
(2) *The Hamiltonian vector field associated to a function $f \in C^\infty(M)$ on a Batalin-Vilkovisky manifold is given by the formula $H_f = -[\Delta, f] + \Delta f$.*
(3) *If $S$ is a function on $M$ and $\Delta_S$ is the operator associated to the Berezinian $\exp(S)\mu$, then $\Delta_S = \Delta - H_S$ and $\Delta_S^2 = H_{\Delta S + \frac{1}{2}[S,S]}$.*

*Proof.* (1) To prove that $C^\infty(M)$ is a Batalin-Vilkovisky algebra, it suffices to verify the formula

$$\Delta(fg) = (\Delta f)g + (-1)^{|f|} f(\Delta g) + (-1)^{|f|}[f, g].$$

This follows by the calculation

$$\begin{aligned}
\operatorname{div}_\mu H_{fg} &= \operatorname{div}_\mu(H_f g + (-1)^{|f|} f H_g) \\
&= (\operatorname{div}_\mu H_f)g - H_f(g) + (-1)^{|f|} f(\operatorname{div}_\mu H_g) - (-1)^{|f|(|g|-1)+|f|} H_g(f) \\
&= (\operatorname{div}_\mu H_f)g + (-1)^{|f|} f(\operatorname{div}_\mu H_g) + 2(-1)^{|f|}[f,g].
\end{aligned}$$

(2) If $f$ and $g$ are functions on $M$, we have

$$\begin{aligned}
[[\Delta, f], g] &= (\Delta f - (-1)^{|f|} f\Delta)g - (-1)^{(|f|+1)|g|} g(\Delta f - (-1)^{|f|} f\Delta) \\
&= \Delta(fg) - (-1)^{|f|} f(\Delta g) - (-1)^{(|f|+1)|g|} g(\Delta f) \\
&= (-1)^{|f|}[f,g] = -H_f(g).
\end{aligned}$$

Thus, $[\Delta, f] = -H_f + \Phi$, where $\Phi$ is a function on $M$. Applying both sides to the function 1, we see that $\Phi = \Delta f$. (3) The formula for $\Delta_S$ is shown as follows:

$$\Delta_S f = \operatorname{div}_{\exp(S)\mu} H_f = \operatorname{div}_\mu H_f + H_f(S) = \Delta f - H_S f.$$

Finally, we obtain the formula for $\Delta_S^2$:

$$\begin{aligned}
\Delta_S^2 f &= (\Delta - H_S)^2 f \\
&= \Delta^2 f - H_S(\Delta f) - \Delta(H_S f) + H_S^2 f \\
&= [S, \Delta f] + \Delta[S, f] + \tfrac{1}{2}[[S, S], f] \\
&= \big[\Delta S + \tfrac{1}{2}[S, S], f\big]. \quad \square
\end{aligned}$$

The equation $\Delta_S^2 = 0$, or $\Delta S + \tfrac{1}{2}[S, S] = 0$, is called the Batalin-Vilkovisky master equation.

The following definition is motivated by the example of the integral $\int_\mu f$ on a Batalin-Vilkovisky supermanifold.

**Definition 2.4.** A trace on a Batalin-Vilkovisky algebra $A$ is a linear form $\operatorname{Tr}: A \to \mathbb{C}$ such that for all $x, y \in A$,

$$\operatorname{Tr}\big((\Delta x)y + (-1)^{|x|-1} x(\Delta y)\big) = \operatorname{Tr}(\Delta x) = 0.$$

Observe that if $A$ is a unital Batalin-Vilkovisky algebra, the vanishing of $\operatorname{Tr}(\Delta x) = 0$ follows from the "integration by parts" formula for $\Delta$, by setting $y = 1$.

**Proposition 2.5.** *If $M$ is a Batalin-Vilkovisky manifold, the integral*

$$f \in C_c^\infty(M) \mapsto \int_\mu f$$

*is a trace on the Batalin-Vilkovisky algebra $C_c^\infty(M)$.*

*Proof.* By the definition of $\Delta f$, we see that

$$\int_\mu (\Delta f)g = \int_\mu (\operatorname{div}_\mu H_f)g = -\int_\mu H_f(g) = (-1)^{|f|} \int_\mu [f,g].$$

The formula $\operatorname{Tr}((\Delta f)g) = (-1)^{|f|} \operatorname{Tr}(f(\Delta g))$ follows from the antisymmetry

$$[f,g] + (-1)^{(|f|-1)(|g|-1)}[g,f] = 0. \quad \square$$

Observe that what we call a Batalin-Vilkovisky algebra bears the same relation to Batalin's and Vilkovisky's theory that Poisson algebras bear to symplectic geometry. For this reason, the name Batalin-Vilkovisky-Poisson algebra might be more appropriate: however, it is a little unwieldy.

## 3. Topological field theory in two dimensions

In this section, we present Segal's definition of a topological conformal field theory in two dimensions, modelled on his definition of a conformal field theory [21]. (This is well-known to physicists under the name of the operator formalism.)

From the spaces $\widehat{\mathcal{M}}_{g,n}$ of the introduction, Segal has constructed a category $\widehat{\mathcal{M}}$ (without identity morphisms). The objects of $\widehat{\mathcal{M}}$ are the natural numbers, while the space of morphisms $\widehat{\mathcal{M}}(m,n)$ is the moduli space of (possibly disconnected) Riemann surfaces of arbitrary genus, together with a biholomorphic map from the disjoint union $\coprod_{i=1}^{m+n} D$ of $m+n$ discs, such that the images of the interiors of the discs are disjoint. The product of two morphisms $\Sigma_1 \in \widehat{\mathcal{M}}(\ell,m)$ and $\Sigma_2 \in \widehat{\mathcal{M}}(m,n)$ is obtained by sewing along the boundaries of the last $m$ discs of $\Sigma_1$ and the first $m$ discs of $\Sigma_2$.

The space of morphisms $\widehat{\mathcal{M}}(m,n)$ is a complex manifold, and the composition map is holomorphic. The space $\widehat{\mathcal{M}}(m,n)$ has an infinite number of components, each of which is a product of moduli spaces $\widehat{\mathcal{M}}_{g,k}$ for some $g$ and $k$:

$$\widehat{\mathcal{M}}(m,n) = \coprod_{i=1}^{\infty} \coprod_{\substack{\pi \in \Pi(m,k) \\ \rho \in \Pi(n,k)}} \coprod_{g_1\ldots g_k=0}^{\infty} \widehat{\mathcal{M}}_{g_1,|\pi_1|+|\rho_1|} \times \cdots \times \widehat{\mathcal{M}}_{g_i,|\pi_k|+|\rho_k|},$$

where $\Pi(m,k)$ is the set of partitions of $\{1,\ldots,m\}$ into $k$ disjoint subsets.

Since the symmetric group $\mathbb{S}_n$ acts freely on $\widehat{\mathcal{M}}_{g,n}$, we see that the group $\mathbb{S}_m^{\operatorname{op}} \times \mathbb{S}_n$ acts freely on $\widehat{\mathcal{M}}(m,n)$, and the composition in $\widehat{\mathcal{M}}$ is equivariant with respect to these actions: composition descends to a map of the quotient

$$\circ : \widehat{\mathcal{M}}(\ell,m) \times_{\mathbb{S}_m} \widehat{\mathcal{M}}(m,n) \to \widehat{\mathcal{M}}(\ell,n).$$

The category $\widehat{\mathcal{M}}$ is a symmetric strictly monoidal category, with tensor product

$$\otimes : \widehat{\mathcal{M}}(m_1,n_1) \times \widehat{\mathcal{M}}(m_2,n_2) \to \widehat{\mathcal{M}}(m_1+m_2, n_1+n_2)$$

induced by disjoint union. This tensor product is equivariant with respect to the symmetric group actions on $\widehat{\mathcal{M}}$.

Let $\chi : \widehat{\mathcal{M}}(m,n) \to \mathbb{Z}$ be the locally constant function equal to the Euler characteristic of the underlying closed surface: this equals $2\sum_i (1-g_i)$, where $g_i$ is the genus of the $i$-th component of the surface. The locally constant function $n - \chi/2 : \widehat{\mathcal{M}} \to \mathbb{Z}$ is additive under both composition and tensor product.

**Definition 3.1.** A topological conformal field theory of ghost number anomaly $d$ consists of the following data:

(1) a bicomplex
$$\mathcal{E} = \sum_{i,j} \mathcal{E}_{i,j},$$
with differentials $Q : \mathcal{E}_{i,j} \to \mathcal{E}_{i-1,j}$ and $\bar{Q} : \mathcal{E}_{i,j} \to \mathcal{E}_{i,j-1}$ (the BRST operators of the chiral and anti-chiral sectors);

(2) a collection of differential forms (the functional integrals),
$$\omega(m,n) \in \Omega^{\bullet,\bullet}(\widehat{\mathcal{M}}(m,n), \mathrm{Hom}(\mathcal{E}^{(m)}, \mathcal{E}^{(n)}))$$
on the complex manifolds $\widehat{\mathcal{M}}(m,n)$ of bidegree $(d(n-\chi/2), d(n-\chi/2))$.

These data are required to satisfy the following conditions:

**(cycle)** $\omega(m,n)$ is closed under the differential $d = (Q + \bar{Q}) = (\partial + Q) + (\bar{\partial} + \bar{Q})$;

**(equivariance)** $\omega(m,n)$ is invariant under the action of $\mathbb{S}_m^{\mathrm{op}} \times \mathbb{S}_n$ on
$$\Omega^{\bullet}(\widehat{\mathcal{M}}(m,n), \mathrm{Hom}(\mathcal{E}^{(m)}, \mathcal{E}^{(n)}))$$
induced by its actions on $\widehat{\mathcal{M}}(m,n)$ and $\mathrm{Hom}(\mathcal{E}^{(m)}, \mathcal{E}^{(n)})$;

**(sewing)** we have the equation $\omega(\ell,m) \circ \omega(m,n) = f^*\omega(\ell,n)$, where
$$\omega(\ell,m) \circ \omega(m,n) \in \Omega^{\bullet}(\widehat{\mathcal{M}}(\ell,m) \times \widehat{\mathcal{M}}(m,n), \mathrm{Hom}(\mathcal{E}^{(\ell)}, \mathcal{E}^{(n)}))$$
denotes the result of applying to the external product $\omega(\ell,m) \boxtimes \omega(m,n)$ the composition
$$\mathrm{Hom}(\mathcal{E}^{(\ell)}, \mathcal{E}^{(m)}) \otimes \mathrm{Hom}(\mathcal{E}^{(m)}, \mathcal{E}^{(n)}) \to \mathrm{Hom}(\mathcal{E}^{(\ell)}, \mathcal{E}^{(n)}),$$
and $f$ is the composition map $f : \widehat{\mathcal{M}}(\ell,m) \times \widehat{\mathcal{M}}(m,n) \to \widehat{\mathcal{M}}(\ell,n)$;

**(disjoint union)** we have the equation
$$\omega(m_1,n_1) \otimes \omega(m_2,n_2) = f^*\omega(m_1+m_2, n_1+n_2),$$
where
$$\omega(m_1,n_1) \otimes \omega(m_2,n_2) \in \Omega^{\bullet}(\widehat{\mathcal{M}}(m_1,n_1) \times \widehat{\mathcal{M}}(m_2,n_2), \mathrm{Hom}(\mathcal{E}^{(m_1+m_2)}, \mathcal{E}^{(n_1+n_2)}))$$
is the external tensor product and $f$ is the disjoint union
$$f : \widehat{\mathcal{M}}(m_1,n_1) \times \widehat{\mathcal{M}}(m_2,n_2) \to \widehat{\mathcal{M}}(m_1+m_2, n_1+n_2).$$

We have chosen to be a little vague about the notion of vector space which we employ here: the vector space $\mathcal{E}$ underlying a topological field theory is really a topological vector space, and the tensor product should be replaced by a suitable topological tensor product.

Unlike in a general conformal field theory, there is no need in the above definition for a line bundle on $\widehat{\mathcal{M}}$ corresponding to a central extension of the Virasoro algebra; this is because topological conformal field theories of necessity have zero conformal anomaly.

We will denote by $\mathcal{H}$ the homology of the graded vector space $\mathcal{E}$ with respect to the BRST differential $Q + \bar{Q}$. There is a non-degenerate quadratic form on $\mathcal{H}$, analogous to the intersection form of a compact oriented manifold

$$\langle \alpha, \beta \rangle = \int_M \alpha \wedge \beta.$$

The quadratic form $\langle v, w \rangle$ corresponds to any connected morphism in $\widehat{\mathcal{M}}(2,0)$ of genus zero, and it is seen to be non-degenerate since it has an inverse in $\mathcal{H} \otimes \mathcal{H}$, induced by a connected morphism in $\widehat{\mathcal{M}}(0,2)$ of genus zero. (For a discussion of this quadratic form, see Section 3 of Witten [23].) The ghost number anomaly $d$ in the definition of a topological conformal field theory manifests itself in the fact that this quadratic form induces a duality between $\mathcal{H}_{i,j}$ and $\mathcal{H}_{-d-i,-d-j}$.

The results of the next section have a generalization to general topological field theories in two dimensions. Let $\mathcal{N}$ be the differentiable category with the same objects as $\widehat{\mathcal{M}}$, namely the natural numbers, and whose space of morphisms $\mathcal{N}(m,n)$ is the disjoint union over all surfaces $\Sigma$ whose boundary $\partial \Sigma$ is a union of $m+n$ circles, of the space of Riemannian metrics such that a tubular neighbourhood of the boundary is isometric to the cylinder $\partial \Sigma \times [0, \varepsilon)$, together with a basepoint on each component of the boundary, modulo the diffeomorphisms of $\Sigma$ which preserve the base-points. Once more, the composition is defined by sewing along the circles, and the symmetric monoidal structure is defined by disjoint union. Two-dimensional topological field theories are defined using the category $\mathcal{N}$ in the same way as topological conformal field theories are defined using $\widehat{\mathcal{M}}$, except that the complex $\mathcal{E}$ is only supposed to be $\mathbb{Z}/2$-graded, and the differential forms $\omega_{m,n}$ are supposed to be of even total degree.

The category $\mathcal{N}$ is fibred over $\widehat{\mathcal{M}}$ by the map which associates to a Riemannian metric on a surface the associated conformal structure: indeed, these two categories are homotopy equivalent. Thus, many of the results which we prove for topological conformal field theories extend to topological field theories in two-dimensions, provided the $\mathbb{Z}$-grading is everywhere replaced by $\mathbb{Z}/2$-grading.

## 4. The Batalin-Vilkovisky operad

An important tool in studying categories of algebras is the notion of a triple. If $\mathcal{C}$ is a category, then the functors $\mathrm{End}(\mathcal{C})$ from $\mathcal{C}$ to itself themselves form a category, with morphisms the natural transformations. This category is a monoidal category, with tensor product the composition $S \circ T$ of functors, and unit the identity functor Id.

**Definition 4.1.** A triple on a category $\mathcal{C}$ is a monoid in the monoidal category $\mathrm{End}(\mathcal{C})$.

More explicitly, a triple is a functor $T : \mathcal{C} \to \mathcal{C}$, together with natural transformations $\mu : TT \to T$ and $\eta : \mathrm{Id} \to T$, such that the diagram

$$\begin{array}{ccc} TTT & \xrightarrow{T\mu} & TT \\ {\scriptstyle \mu T}\downarrow & & \downarrow{\scriptstyle \mu} \\ TT & \xrightarrow{\mu} & T \end{array}$$

commutes, and the two compositions

$$T \xrightarrow{T\eta} TT \xrightarrow{\mu} T$$
$$T \xrightarrow{\eta T} TT \xrightarrow{\mu} T$$

are the identity natural transformation.

Suppose that $\mathcal{C}$ is a symmetric monoidal category with colimits (fibred sums). Define an $\mathbb{S}$-module in $\mathcal{C}$ to be a sequence $k \mapsto \mathbf{a}(k)$ of representations of the symmetric groups $\mathbb{S}_k$ in $\mathcal{C}$. To an $\mathbb{S}$-module $\mathbf{a}$, we may associate an analytic functor $T(\mathbf{a}) : \mathcal{C} \to \mathcal{C}$ by the formula

$$T(\mathbf{a}, V) = \sum_{k=0}^{\infty} \mathbf{a}(k) \otimes_{\mathbb{S}_k} V^{(k)}.$$

Analytic functors are a particularly explicit type of functor, which may be studied by means of their "Taylor coefficients," the $\mathbb{S}$-module $\mathbf{a}$.

The notion of an operad evolved through the work of Boardman and Vogt [4] and May [17]. Their definition is equivalent to the following one.

**Definition 4.2.** An operad in a symmetric monoidal category $\mathcal{C}$ is an $\mathbb{S}$-module $\mathbf{a}$ together with the structure of a triple on the analytic functor $T(\mathbf{a})$.

Thus, an operad is nothing other than the Taylor coefficients of an analytic triple. The structure of an operad on an $\mathbb{S}$-module $\mathbf{a}$ is determined by composition maps

$$\mathbf{a}(k) \times \mathbf{a}(j_1) \times \cdots \times \mathbf{a}(j_k) \to \mathbf{a}(j_1 + \cdots + j_k),$$

which satisfy conditions of equivariance and associativity explained in May [17] (see also [10]).

We will be interested in operads in the symmetric monoidal category of differentiable manifolds with tensor product the product, and in the symmetric monoidal category of graded vector spaces; we call these repsectively differentiable and linear operads. Given a differentiable operad $\mathcal{M}$, its homology $\mathbf{a}(k) = H_\bullet(\mathcal{M}(k))$ is a linear operad.

Associated to any triple is its category of algebras $\mathcal{C}^T$.

**Definition 4.3.** An algebra over a triple $(T, \mu, \eta)$ is a pair $(V, \rho)$ where $V$ is an object of the category $\mathcal{C}$ and $\rho : TV \to V$ is a morphism, such that the composition

$$V \xrightarrow{\eta V} TV \xrightarrow{\rho} V$$

is the identity, and the following diagram commutes:

$$\begin{array}{ccc} TTV & \xrightarrow{T\rho} & TV \\ {\mu V}\downarrow & & \downarrow{\rho} \\ TV & \xrightarrow{\rho} & V \end{array}$$

A morphism $\phi : V_1 \to V_2$ of $\mathcal{C}$ is a map of algebras if the diagram

$$\begin{array}{ccc} TV_1 & \xrightarrow{T\phi} & TV_2 \\ {\rho_1}\downarrow & & \downarrow{\rho_2} \\ V_1 & \xrightarrow{\phi} & V_2 \end{array}$$

commutes. The category of $T$-algebras is denoted $\mathcal{C}^T$.

By way of illustration, let us describe three examples of linear operads.

(1) Let $A$ be a graded vector space and let $T_A$ be the functor $V \mapsto A \otimes V$; this is an analytic functor, corresponding to the $\mathbb{S}$-module $\mathbf{a}$ with $\mathbf{a}(1) = A$ and $\mathbf{a}(k) = 0$ for $k \neq 1$. Then the structure of a triple on $T_A$ corresponds to the structure of an algebra on $A$, and an algebra for the triple $T_A$ is a left $A$-module.
(2) Let $T$ be the tensor functor

$$TV = \sum_{k=0}^{\infty} V^{(k)};$$

this is an analytic functor, corresponding to the $\mathbb{S}$-module $\mathbf{a}(k) = \mathbb{C}[\mathbb{S}_k]$. Then $T$ is a triple, with product $\mu : TTV \to TV$ defined by mapping

$$(v_{1,1} \otimes \ldots \otimes v_{1,k_1}) \otimes \ldots \otimes (v_{n,1} \otimes \ldots \otimes v_{n,k_n}) \in TTV$$

to $v_{1,1} \otimes \ldots \otimes v_{1,k_1} \otimes \ldots \otimes v_{n,1} \otimes \ldots \otimes v_{n,k_n} \in TV$, and with unit $\eta : V \to TV$ defined by the inclusion of $V$ as a summand in $TV$. An algebra over the triple $T$ is the same thing as a graded associative algebra.

(3) Similarly, the symmetric tensor functor

$$SV = \sum_{k=0}^{\infty} (V^{(k)})_{\mathbb{S}_k},$$

associated to the $\mathbb{S}$-module $\mathbf{c}(k) = \mathbb{C}$ the trivial representation of $\mathbb{S}_k$, is a triple, and an algebra over $S$ is the same thing as a graded commutative algebra. (Here $V_{\mathbb{S}_k}$ denotes the space of coinvariants $\{v - gv \mid v \in V, g \in G\}$ of an $\mathbb{S}_k$-module $V$.)

We now define operads whose categories of algebras are the category of braid algebras and the category of Batalin-Vilkovisky algebras.

Given a Coxeter group $(W, S)$ generated by involutions $s_i \in S$ satisfying the braid relations

$$\underbrace{s_i s_j \ldots}_{d_{ij} \text{ times}} = \underbrace{s_j s_i \ldots}_{d_{ij} \text{ times}},$$

there is a generalized braid group $\mathbb{B}_W$, defined by generators $\sigma_i$ which satisfy the braid relations, and a surjective map $\mathbb{B}_W \to W$, defined by sending $\sigma_i$ to $s_i$.

The braid group $\mathbb{B}_k$ is the braid group associated to the Coxeter group $A_{k-1}$, which is isomorphic to the symmetric group $\mathbb{S}_k$, and has Coxeter matrix $(d_{ij})_{1 \leq i < j \leq k-1}$

$$d_{ij} = \begin{cases} 3, & |i-j| = 1 \\ 2, & |i-j| > 1. \end{cases}$$

Define the pure braid group $\mathbb{P}_k$ on $k$ strands by the short exact sequence

$$1 \to \mathbb{P}_k \to \mathbb{B}_k \to \mathbb{S}_k \to 1.$$

Let $\mathbf{b}$ be the $\mathbb{S}$-module $\mathbf{b}(k) = H_\bullet(\mathbb{P}_k)$; the associated analytic functor $T(\mathbf{b})$ may be written

$$T(\mathbf{b}, V) = \sum_{k=0}^{\infty} H_\bullet(\mathbb{S}_k, \mathbf{b}(k) \otimes V^{(k)}) = \sum_{k=0}^{\infty} H_\bullet(\mathbb{B}_k, V^{(k)}),$$

where $\mathbb{B}_k$ acts on $V^{(k)}$ via its homomorphism to the symmetric group $\mathbb{S}_k$.

The Coxeter group $B_k$, with Coxeter matrix $(d_{ij})_{1 \leq i < j \leq k}$

$$d_{ij} = \begin{cases} 4, & |i-j| = 1, j = k, \\ 3, & |i-j| = 1, j < k, \\ 2, & |i-j| > 1, \end{cases}$$

is isomorphic to the hyperoctahedral group $\mathbb{S}_k \wr (\mathbb{Z}/2) \cong (\mathbb{Z}/2)^k \rtimes \mathbb{S}_k$. The corresponding generalized braid group is isomorphic to $\mathbb{B}_k \wr \mathbb{Z} \cong \mathbb{Z}^k \rtimes \mathbb{B}_k$. As is explained in Chapter 4 of [14], this group may be realized as the group of braids on $k$ ribbons: the generator $\sigma_i$, $1 \leq i \leq k-1$, braids the $i$-th ribbon beneath the $i+1$-th ribbon, while $\sigma_k$ takes the $k$-th ribbon through a half-twist. We now define the analogue for the ribbon braid groups of the functor $T_{\mathbf{b}}$: if $\mathbf{bv}$ is the $\mathbb{S}$-module $\mathbf{bv}(k) = H_\bullet(\mathbb{Z}^k \times \mathbb{P}_k)$, the associated analytic functor is

$$T(\mathbf{bv}, V) = \sum_{k=0}^\infty H_\bullet(\mathbb{S}_k, \mathbf{bv}(k) \otimes V^{(k)}) = \sum_{k=0}^\infty H_\bullet(\mathbb{B}_k \wr \mathbb{Z}, V^{(k)}).$$

We will show that $T(\mathbf{b})$ and $T(\mathbf{bv})$ are the underlying functors of triples.

**Theorem 4.4.** *There is natural equivalence between the category of algebras for the triple $T(\mathbf{b})$ and the category of unital braid algebras.*

**Theorem 4.5.** *There is a natural equivalence between the category of algebras for the triple $T(\mathbf{bv})$ and the category of unital Batalin-Vilkovisky algebras.*

Theorem 4.4 is due to F. Cohen [5] (for a different proof of this result, see [10]), while Theorem 4.5 is, to the best of our knowledge, new.

One way of interpreting Theorem 4.4 is that there is a natural correspondence between elements of $H_\bullet(\mathbb{P}_k)$ and expressions that may be formed from $k$ elements $(a_1, \ldots, a_k)$ of a braid algebra using the product of the bracket, and such that each element $a_i$ occurs just once. Since braid algebras are relatively unfamiliar, let us illustrate this correspondence for $k \leq 3$.

($k = 1$) The only word in a single element $x$ is $x$ itself, in degree 0. This corresponds to the fact that $\mathbf{b}(1) \cong \mathbb{C}$.

($k = 2$) The possible words in elements $x$ and $y$ are $xy$, in degree 0, and $(-1)^{|x|}[x, y]$, in degree 1. The group $\mathbb{S}_2$ acts trivially on both of these. We see that $\mathbf{b}(2)$ may be naturally identified to the homology of a circle, corresponding to the fact that $\mathbb{P}_2 \cong \mathbb{Z}$.

($k = 3$) The possible words in elements $x$, $y$ and $z$ are

$$\begin{aligned}
\text{degree} &= 0: & & xyz \\
\text{degree} &= 1: & [x,y]z \quad & [y,z]x \quad [z,x]y \\
\text{degree} &= 2: & [[x,y],z] \quad & \quad [[y,z],x]
\end{aligned}$$

Indeed, the Poincaré polynomial of $\mathbf{b}(3)$ equals $(1 + t)(1 + 2t) = 1 + 3t + 2t^2$.

To show that $T(\mathbf{b})$ and $T(\mathbf{bv})$ are triples, it suffices to show that $\mathbf{b}(k) = H_\bullet(\mathbb{P}_k)$ and $\mathbf{bv}(k) = H_\bullet(\mathbb{Z}^k \times \mathbb{P}_k)$ are linear operads. We do this by constructing differentiable operads of which they are the homology.

Recall the "little disc operad" $\mathcal{C}$ of Boardman and Vogt [4] (see also May [17]). Let $D$ be the unit disc in the complex plane, and let $\mathcal{C}(k)$ be the space of all maps from $\coprod_{i=1}^k D$ to $D$ which when restricted to each disc is the composition of a translation and a dilation, and such that the images of the discs are disjoint. The operad structure is given by the maps

$$\mathcal{C}(k) \times \mathcal{C}(j_1) \times \cdots \times \mathcal{C}(j_k) \to \mathcal{C}(j_1 + \cdots + j_k),$$

where for $(c, d_1, \ldots, d_k) \in \mathcal{C}(k) \times \mathcal{C}(j_1) \times \cdots \times \mathcal{C}(j_k)$, the element $c(d_1, \ldots, d_k)$ operates on $\ell$-th disc $D$, $j_i + 1 \leq \ell \leq j_{i+1}$, by the action of $d_i$ on the $(\ell - j_i)$-th disc followed by the action of $c$ on the $i$-th disc. Note that the spaces $\mathcal{C}(k)$ are open subsets of $\mathbb{C}^{2k}$.

The map from $\mathcal{C}(k)$ to the configuration space
$$F_k(D) = \{(x_1, \ldots, x_k) \mid x_i \neq x_j \text{ for } i \neq j\}$$
defined by sending a map from $D$ to itself to the image of its centre, is a a fibration with contractible fibres, hence a homotopy equivalence (May [17], page 34). It may be shown by induction on $k$ (Fadell and Neuwirth [7]) that $F_k(D) \simeq K(\mathbb{P}_k, 1)$, This gives the $\mathbb{S}$-module $\mathbf{b}(k) \cong H_\bullet(\mathcal{C}(k))$ the structure of a linear operad.

Arnold [1] has calculated the homology of the operad $\mathcal{C}(k)$, or rather, of the configuration spaces $F_k(\mathbb{C})$.

**Proposition 4.6.** *For $1 \leq i \neq j \leq k$, let $\omega_{ij} \in H^1(F_k(\mathbb{C}), \mathbb{Z})$ be the inverse image of the generator of $H^1(\mathbb{C}^\times, \mathbb{Z})$ under the map $F_k(\mathbb{C}) \xrightarrow{z_i - z_j} \mathbb{C}^\times$. The homology $H_\bullet(F_k(\mathbb{C}), \mathbb{Z})$ is torsion free, and its dual $H^\bullet(F_k(\mathbb{C}), \mathbb{Z})$ is the graded commutative ring with generators $\omega_{ij} \in H^1(F_k(\mathbb{C}), \mathbb{Z})$, and relations*

(1) $\omega_{ji} = \omega_{ij}$;
(2) $\omega_{ij}\omega_{jk} + \omega_{jk}\omega_{ki} + \omega_{ki}\omega_{ij} = 0$.

*The symmetric group $\mathbb{S}_k$ acts on $H^\bullet(F_k(\mathbb{C}), \mathbb{Z})$ by $\pi^* \omega_{ij} = \omega_{\pi(i)\pi(j)}$.*

As a corollary, we see that the Poincaré polynomial of $\mathbf{b}(k)$ equals
$$\sum_{i=0}^\infty \dim H_i(\mathbb{P}_k) t^i = (1 + t)(1 + 2t) \ldots (1 + (k-1)t).$$

In fact, Arnold shows that the cohomology class $\omega_{ij}$ is represented in de Rham cohomology by the closed one-form
$$\omega_{ij} = \frac{1}{2\pi i} d \log(z_i - z_j),$$
and the differential forms $\omega_{ij}$ already satisfy the above relations. The cohomology class $\omega_{ij}$ is dual to the image under the Hurewicz map of the element $A_{ij} \in \mathbb{P}_k$ given by the formula
$$A_{ij} = \sigma_{j-1}\sigma_{j-2} \ldots \sigma_{i+1} \sigma_i^2 \sigma_{i+1}^{-1} \ldots \sigma_{j-2}^{-1} \sigma_{j-1}^{-1}.$$
For a picture of this braid, see page 21 of Birman [3].

Our proof of Theorem 4.4 in [10] proceeds by showing that $\mathbf{b}(k)$ is quasi-isomorphic to a chain complex with summands indexed by trees with $k$ leaves. These trees correspond to strata of the Fulton-MacPherson compactification of the configuration space $F_k(\mathbb{C})$ [8] (or rather, its analogue in the category of differentiable manifolds with corners). The theorem follows from the collapse of the spectral sequence for the homology of this stratified space at its $E^2$-term, which may be shown either by the use of Deligne's mixed Hodge theory, as in Beilinson-Ginzburg [2], or by an argument which compares the spectral sequences for the Fulton-MacPherson compactifications of $F_k(\mathbb{R}^n)$ as $n$ varies, and makes use of the generalization of Proposition 4.6 to $F_k(\mathbb{R}^n)$, $n > 1$: the classes $\omega_{ij}$ are then in $H^{n-1}(F_k(\mathbb{R}^n), \mathbb{Z})$ (see [5]).

A variant of the little disc operad is the little framed disc operad, defined as follows: $\mathcal{P}(k)$ is the space of all maps from $\coprod_{i=1}^k D$ to $D$ which restrict on each disc to a composition of a translation and multiplication by an element of $\mathbb{C}^\times$, and such that the images of the discs are disjoint. The operad structure is defined by composition, in the same way as for the little disc operad $\mathcal{C}$. There is an equivariant map from $\mathcal{P}(k)$ to $\mathcal{C}(k)$, defined by forgetting the angle of a disc: this map is a fibration with fibre the torus $\mathbb{T}^k$. This shows that the $\mathbb{S}$-module $\mathbf{bv}(k) = H_\bullet(\mathbb{Z}^k \times \mathbb{P}_k)$ is an operad.

Associated to the category $\widehat{\mathcal{M}}$ of the last section is an operad $\widehat{\mathcal{M}}_0$, with $\widehat{\mathcal{M}}_0 = \widehat{\mathcal{M}}_{0,k+1} \subset \widehat{\mathcal{M}}(k,1)$. There is an inclusion of operads $\mathcal{P} \to \widehat{\mathcal{M}}_0$, defined by considering an element of $\mathcal{P}(k)$ to be a special kind of element of $\widehat{\mathcal{M}}_{0,k+1}$: we identify the range $D$ in the definition of an element of $\mathcal{P}(k)$ with the unit disc in $S^2$, and map the zeroth disc to $S^2$ by the disc at infinity by the map $z \mapsto -1/z$. The map $\mathcal{P}(k) \to \widehat{\mathcal{M}}_0(k) = \widehat{\mathcal{M}}_{0,k+1}$ is a homotopy equivalence, showing that $\mathbf{bv}(k) = H_\bullet(\widehat{\mathcal{M}}_{0,k+1})$.

The restriction of a conformal field theory to the operad $\widehat{\mathcal{M}}_0$ may be thought of as its classical limit, since $\widehat{\mathcal{M}}_0$ corresponds to Feynman diagrams with the topology of a tree. Note that on $\widehat{\mathcal{M}}_0$, the integer $d(n - \chi/2)$ equals $0$: thus, the differential forms $\omega(m,n)$ when restricted to $\widehat{\mathcal{M}}_0(m,n)$ have total degree $(0,0)$.

Suppose $\mathcal{E}$ is the chain complex underlying a topological conformal field theory, with homology $\mathcal{H} = H_\bullet(\mathcal{E}, Q + \bar{Q})$. Given a cycle $\gamma$ in $\widehat{\mathcal{M}}_0(k)$, we may integrate $\omega(k,1)$ over $\gamma$ to obtain a closed operator from $\mathcal{E}^{(k)}$ to $\mathcal{E}$, of degree equal to the degree of $\gamma$. The operator from $\mathcal{H}^{(k)}$ to $\mathcal{H}$ induced by this operator does not depend on the homology class of $\gamma$. In this way, we see that $\mathcal{H}$ is an algebra for the triple $T(\mathbf{bv})$, and hence by Theorem 4.5 is a unital Batalin-Vilkovisky algebra.

Denote by $\int$ the linear form on $\mathcal{H}$ defined by a morphism in $\widehat{\mathcal{M}}(1,0)$ of genus 0: this is just the operation of sewing a disk along a boundary circle.

**Proposition 4.7.** *The linear form $\int : \mathcal{H} \to \mathbb{C}$ is a trace on the Batalin-Vilkovisky algebra $\mathcal{H}$. The non-degenerate quadratic form $\langle v, w \rangle$ on $\mathcal{H}$ is given by the formula*

$$\langle v, w \rangle = \int vw.$$

*Proof.* The formula relating $\langle v, w \rangle$ and $\int$ is follows from the fact that the cylinder is obtained from a morphism of $\widehat{\mathcal{M}}_0(2,1)$ (a pair of pants) by composing with the disc $D$ in $\widehat{\mathcal{M}}(1,0)$.

To show that $\int$ is a trace, we must show that for all $v, w \in \mathcal{H}$,

$$\int (\Delta v) w = (-1)^{|v|} \int v(\Delta w).$$

This formula follows from the fact that $\mathcal{M}(2,0)$ is homotopy equivalent to a circle: thus, the two sides are *a priori* linearly dependent, and the constant is easily found. $\square$

If the topological conformal field theory associated to $\mathcal{E}$ has ghost number anomaly $d$, the resulting trace $\int$ is concentrated in bidegree $(-d, -d)$.

We may also show that the homology $H_\bullet(\Omega^2 M)$ of a double loop space is an algebra for the triple $T(\mathbf{bv})$. (Haynes Miller has pointed out that this reflects the fact that the H-space of based self-maps of the two-sphere $S^2$ homotopic to the identity is rationally equivalent to the circle group $S^1$.) This is shown using maps

$$\mathcal{P}(k) \times (\Omega^2 M)^k \to \Omega^2 M$$

which generalize those of Boardman-Vogt and May: given $c = (z_1, \ldots, z_k) \in \mathcal{P}(k)$ and elements $(f_1, \ldots, f_k) \in (\Omega^2 M)^k$, we define $c(f_1, \ldots, f_k) \in \Omega^2 M$ to be the double loop which sends $z \in D$ to $f_i(z_i^{-1}(z)) \in M$ if $z_i$ lies in the image of the $i$-th disc, and to the basepoint of $X$ otherwise.

We finally return to the proof of Theorem 4.5. This is easy granted the corresponding result for braid algebras. First, we give a description of $\mathbf{bv}(k) = H_\bullet(\mathcal{P}(k))$ of the same type as the description of $\mathbf{b}(k)$ given by Arnold [1].

**Proposition 4.8.** *For $1 \leq i \neq j \leq k$, let $\omega_{ij} \in H^1(\mathcal{P}(k), \mathbb{Z})$ be the inverse image of the generator of $H^1(\mathbb{C}^\times, \mathbb{Z})$ under the map $\mathcal{P}(k) \xrightarrow{z_i(0) - z_j(0)} \mathbb{C}^\times$. For $1 \leq i \leq k$, let $\eta_i \in H^1(\mathcal{P}(k), \mathbb{Z})$ be the inverse image of the generator of $H^1(\mathbb{C}^\times, \mathbb{Z})$ under the map $\mathcal{P}(k) \xrightarrow{z_i'(0)} C^\times$. The cohomology $H^\bullet(\mathcal{P}(k), \mathbb{Z})$ is the graded commutative ring with generators $\omega_{ij}, \eta_i \in H^1(\mathcal{P}(k), \mathbb{Z})$, and relations*

(1) $\omega_{ji} = \omega_{ij}$;
(2) $\omega_{ij}\omega_{jk} + \omega_{jk}\omega_{ki} + \omega_{ki}\omega_{ij} = 0$;
(3) $\eta_i \eta_j = \eta_i \omega_{jk} = 0$.

*The symmetric group $\mathbb{S}_k$ acts on $H^\bullet(\mathcal{P}(k), \mathbb{Z})$ by $\pi^* \omega_{ij} = \omega_{\pi(i)\pi(j)}$ and $\pi^* \alpha_i = \alpha_{\pi(i)}$.*

Denote by $\widetilde{\mathbf{bv}}$ the operad describing Batalin-Vilkovisky algebras; $\widetilde{\mathbf{bv}}(k)$ is the vector space spanned by words in the free Batalin-Vilkovisky algebra generated by $\{x_1, \ldots, x_k\}$ which contain each letter $x_i$ precisely once. We must show that $\mathbf{bv}$ and $\widetilde{\mathbf{bv}}$ are isomorphic as operads. It is easily seen that the basis of $\widetilde{\mathbf{bv}}(k)$ consists of words of the form

$$p(\Delta^{\varepsilon_1} x_1, \ldots, \Delta^{\varepsilon_k} x_k),$$

where $p$ ranges over all elements of $\mathbf{b}(k)$, the operad for braid algebras, and $\varepsilon_i \in \{0, 1\}$. This is because all applications of $\Delta$ may be passed down to the bottom level of a word by repeated application of the formulas

$$\Delta(ab) = (\Delta a)b + (-1)^{|a|}a(\Delta b) + (-1)^{|a|}[a, b],$$
$$\Delta[a, b] = [\Delta a, b] + (-1)^{|a|-1}[a, \Delta b].$$

This shows that as an $\mathbb{S}_k$-module, $\widetilde{\mathbf{bv}}(k) \cong \widetilde{\mathbf{bv}}(1)^{(k)} \otimes \mathbf{b}(k)$. Thus, $\mathbf{bv}$ and $\widetilde{\mathbf{bv}}$ are isomorphic as $\mathbb{S}$-modules.

Since $\mathbf{b}$ is generated by $\mathbf{b}(2)$, it follows that $\mathbf{bv}$ is generated as an operad by $\mathbf{bv}(1)$ and $\mathbf{bv}(2)$. We have already observed that $\widetilde{\mathbf{bv}}$ is generated by $\widetilde{\mathbf{bv}}(1)$ and $\widetilde{\mathbf{bv}}(2)$. Thus, it suffices to check that the relations are the same in both cases, in other words, that the composition

$$\mathbf{bv}(1) \otimes \mathbf{bv}(2) \to \mathbf{bv}(2)$$

correctly describes the action of $\Delta$ on $x_1 x_2$ and $[x_1, x_2]$. The adjoint of this map is the homomorphism of algebras $H^\bullet(\mathcal{P}(2)) \to H^\bullet(\mathcal{P}(1)) \otimes H^\bullet(\mathcal{P}(2))$ defined on generators by

$$\omega \mapsto 1 \otimes \omega + \eta \otimes 1,$$
$$\eta_i \mapsto 1 \otimes \eta_i + \eta \otimes 1.$$

It is easy to read off the needed equations: the first equation implies the action of $\Delta$ on $x_1 x_2$ and $[x_1, x_2]$, while the second shows that $\Delta^2 = 0$.

Department of Mathematics, MIT, Cambridge MA 02139 USA
*E-mail address*: getzler@math.mit.edu